\documentclass[10pt, conference, compsocconf]{IEEEtran}
\newcommand{\yt}{\texttt{yt}}

\def\refjnl#1{{\rmfamily#1}}%
\newcommand\apj{\refjnl{ApJ}}%
\newcommand\apjl{\refjnl{ApJ}}%
\newcommand\apjs{\refjnl{ApJS}}%
\newcommand\aap{\refjnl{A\&A}}%
\newcommand\mnras{\refjnl{MNRAS}}%
\newcommand\nat{\refjnl{Nature}}%
\usepackage{cite}

\ifCLASSINFOpdf
  \usepackage[pdftex]{graphicx}
\else
  \usepackage[dvips]{graphicx}
\fi
\usepackage[caption=false,font=footnotesize]{subfig}
\begin{document}
%
\title{High-Performance Astrophysical Simulations and Analysis with Python}


\author{\IEEEauthorblockN{Matthew J.~Turk}
\IEEEauthorblockA{Department of Astronomy\\
Columbia University\\
New York, NY\\
matthewturk@gmail.com}
\and
\IEEEauthorblockN{Britton D.~Smith}
\IEEEauthorblockA{Department of Physics and Astronomy\\
Michigan State University\\
East Lansing, MI\\ 
brittonsmith@gmail.com}
}


%


\maketitle

\begin{abstract}
The usage of the high-level scripting language Python has enabled new
mechanisms for data interrogation, discovery and visualization of scientific
data.  We present \yt{}\footnote{Available at
\texttt{http://yt-project.org/}}, an open source, community-developed
astrophysical analysis and visualization toolkit for data generated by
high-performance computing (HPC) simulations of astrophysical phenomena.
Through a separation of responsibilities in the underlying Python code, \yt{}
allows data generated by incompatible, and sometimes even directly competing,
astrophysical simulation platforms to be analyzed in a consistent manner,
focusing on physically relevant quantities rather than quantities native to
astrophysical simulation codes.  We present on its mechanisms for data access,
capabilities for MPI-parallel analysis, and its implementation as an \textit{in
situ} analysis and visualization tool.

\end{abstract}


%
\IEEEpeerreviewmaketitle

\section{Introduction}\label{sec:introduction}
In the last decade, multiphysics astrophysical simulations have increased
exponentially in both sophistication and size \cite{
2005Natur.435..629S, 
2008JPhCS.125a2008K, 
2009JCoPh.228.6833R, 
2008MNRAS.390.1326O, 
2007arXiv0705.1556N, 
2010arXiv1008.4368K, 
2010arXiv1008.2801A, 
2010arXiv1002.3660K, 
2000ApJS..131..273F, 
2007ApJ...659L..87A}; 
however, the software tools to mine those simulations have not kept pace.
Typically, methods for examining data suffer from a lack of agility,
discouraging exploratory investigation.  To accommodate this, massively
parallel visualization tools such as VisIT and ParaView \cite{visit_paper,
paraview_paper} have been repurposed as domain-specific astrophysical tools.
This repurposing, while effective, does not benefit from domain-specific
analysis or data structures.  The lack of domain-specific quantitative analysis
tools designed for astrophysical data leads to the development of specialized
tools by individual researchers or research groups, most of which are never
shared outside the research group.  This can substantially inhibit
collaboration between different groups--even those using the same simulation
code.

This fractionation of the astrophysical community demonstrates a clear need for
a flexible and cross-code software package for quantitative data analysis and
visualization.  In this paper we present \yt{} \cite{yt_full_paper}, a data
analysis and visualization package that works with several astrophysical
simulation codes.  \yt{} is developed openly and is freely available at
\texttt{http://yt-project.org/}.  It has been designed to be a common platform
for simulation analysis, so that scripts can be shared across groups and
analysis can be repeated by independent scientists\footnote{A platform for
sharing scripts is provided with yt, with command-line helpers, at
\texttt{http://hub.yt-project.org}.}.  By making this tool available, we hope
not only to encourage cross-group collaboration and validation of results, but
to remove or at least greatly lower the barrier to entry for exploratory
simulation analysis.  \yt{} provides mechanisms for conducting complete
analysis pipelines resulting in publication quality figures and data tables, as
well as the necessary components for constructing new methods for examining
data.  The concepts for data handling and representation in \yt{} are certainly
not new, but their application to astrophysical data enables complex, detailed
analysis pipelines to be shared between individuals studying disparate
phenomena using disparate methods.  This enables and even encourages
reproducibility and independent verification of results.

We have built this analysis and visualization code in Python, using NumPy
\cite{numpy_paper} for fast mathematical operations, \texttt{mpi4py} for
MPI-parallelism \cite{MPI4PY:PAPER1, MPI4PY:PAPER2}, and optionally Matplotlib
for 2D visualization \cite{matplotlib_paper}.  Additionally, several core
library routines in \yt{} such as the AMR volume rendering, multi-dimensional
binning, and file access routines, are written in Cython.  In addition to
utilizing community-developed Python modules, \yt{} is itself a Python module
suitable for direct scripting or access as a library.  A community of users and
developers has grown around the project, with over 20 committers in the history
of the project, and it has been used in numerous published papers and posters.
(See, for example, \cite{ 2010ApJ...715.1575S, 2010ApJ...721.1105B,
2009Sci...325..601T, 2011ApJ...738...54K, 2011arXiv1108.4427Z,
2011ApJ...737...63A, 2011MNRAS.414.2297I, 2011ApJ...735...49M,
2011ApJ...731...59C}.)

In order to accomodate the diverse computing environments on which
astrophysical simulations are run, \yt{} was designed to use primarily
off-screen rendering and scripting interfaces, although several smaller tools
are provided for specific, interactive visualization tasks.  The former method
is well-suited to remote visualization and can be run via a job execution queue
on a batch-compute cluster, such as those on which the underlying simulation
are run.  \yt{} is subdivided into several sub-packages for data handling, data
analysis, and plotting.  This modularity encourages the creation of reusable
components for multi-step analysis operations.

While work continues on the exploratory, post-processing methods of data
analysis and visualization that \yt{} was originally designed for, current
development has focused on analysis during the course of a simulation, or
so-called \textit{in situ} analysis.  This allows for high-cadence analysis to
be conducted without writing data to disk.  Future simulations, such as those
to be conducted on petascale machines, will require analysis and visualization
during the simulation rather than exclusively as a post-processing technique.

In this paper, we will describe the mechanisms that \yt{} provides for
accessing data (\S \ref{sec:mechanisms}), methods of interacting with \yt{} (\S
\ref{sec:interacting}), the visualization techniques offered
by \yt{} (\S \ref{sec:visualization}),  the parallelism strategy for data
analysis and generation of visualizations (\S \ref{sec:parallelism}), and end
with a discussion of the process of embedding \yt{} in running simulation codes
and how this will be inverted in the future (\S \ref{sec:embedding}).

\section{Mechanisms for Interacting with Data}\label{sec:mechanisms}

\yt{} primarily operates on Adaptive Mesh Refinement (AMR) data, where the simulation
domain is divided into spatially-organized zones along a regular mesh.  Regions
requiring higher degrees of refinement to capture shocks, collapse,
instabilities and so on are replaced with higher-resolution meshes; in this
manner, a stable and accurate solution is ensured at all locations.  (For more
information, see one of several astrophysics papers describing AMR implementations, such as
\cite{2000ApJS..131..273F,bryan97,1997ApJS..111...73K}.)  While \yt{} is able
to analyze collisionless particles (such as dark matter or star particles),
it is currently ill-suited to analysis of Smoothed Particle Hydrodynamics (SPH)
simulations; future versions will improve support.

The vast majority of AMR calculations in the astrophysical literature are
computed on a rectilinear grid; while this affords a number of computational
efficiencies and conveniences, astrophysical phenomena as a whole are not
rectangular prisms and thus are poorly suited for analysis as rectangular prisms.
This presents a fundamental disconnect between the data structures utilized by
simulations and the geoemetries found in nature.  Furthermore, the task of
selecting geometric regions in space requires substantial overhead: masking of
overlapping simulation regions, selective IO, data selection, and so on.  \yt{}
provides a number of convenience functions and mechanisms for addressing data
within astrophysical simulations that make the process of handling and
manipulating data straightforward.

The \yt{} codebase has been organized along several conceptual lines, each
corresponding to a set of tasks or classes in Python.  The primary mechanisms
for handling data are contained in the Python module \texttt{yt.data\_objects},
while all code and data structures specific to a particular simulation code
resides within a submodule of \texttt{yt.frontends} (such as
\texttt{yt.frontends.enzo}, \texttt{yt.frontends.orion}, etc).  In the current
version of \yt{} full support is provided for accessing and reading Enzo
\cite{bryan97}, FLASH \cite{2000ApJS..131..273F}, Orion
\cite{2007ApJ...667..626K} and Nyx codes, with preliminary support for RAMSES
\cite{2002A&A...385..337T}, ART \cite{1997ApJS..111...73K} and several others.

To open a dataset, the user creates an instance of a simulation code-specific
subclass of \texttt{StaticOutput}, a lightweight class that scans a parameter
file and obtains the necessary information to orient the dataset: the current
time in the simulation, the domain information, the mechanisms for converting
units, and the necessary file locations on disk.  A convenience function
(\texttt{load}) for automatically creating such an instance is provided, such
that it only requires a path on disk to the dataset of interest.  However,
geometric information about the manner in which data is laid out on disk or in
the simulation domain is compartmentalized to a \texttt{AMRHierarchy} object.
These objects are comparatively expensive to construct, as they contain a
hierarchy of \texttt{GridPatch} objects, all of which posses spatial and
parentage information.  These objects are not instantiated or constructed until
requested.  All data access is mediated by \texttt{AMRHierarchy} objects, as
noted below.

By relegating data handling to individual instances of classes, we
compartmentalize datasets; because each dataset is merely a variable, the
number that can be opened and simultaneously cross-compared is only limited by
the available memory and processing power of the host computer.  Furthermore,
datasets from different simulation codes can be opened and compared
simultaneously in memory.

When handling astrophysical data, it is appropriate to speak of geometric
regions that outline the rough boundaries of physical objects: dark matter
halos as ellipsoids, protostars as spheres, spiral galaxies as cylinders, and
so on.  The central conceit behind \yt{} is the presentation to the user of a
series of physical objects with the underlying simulation largely abstracted.
For AMR data, this means hiding the language of grid patches, files on disk and
their interrelationships, and instead describing only geometric or physical
systems; these intermediate steps are handled exclusively by \yt{}, without
requiring any intervention on the part of the user.  For instance, to select a
spherical region, the user specifies a center and a radius and the underlying
\yt{} machinery will identify grid patches that intersect that spherical
region, identify which grid patches are the most highly-refined at all regions
within the sphere, locate the appropriate data on disk, read it and return this
data to the user.  By abstracting the selection of and access to data in this
manner, operations that can be decomposed spatially or that are
``embarrassingly parallel" can be transparently parallelized, without requiring
the user's intervention.   The data containers implemented in \yt{} include
spheres, rectangular prisms, cylinders (disks), arbitrary regions based on
logical operations, topologically-connected sets of cells, axis-orthogonal and
arbitrary-angle rays, and both axis-orthogonal and arbitrary-angle slices.

Data containers provide several methods for data access.  The data can be
accessed directly, as in the above code listing, or through abstractions such
as \emph{object quantities}, where bulk operations are conducted such as
calculating the angular momentum vector or the total mass.

The abstraction of data into data containers leads to the creation of systems
of components: data containers become ``sources'' for both analysis procedures
as well as visualization tasks.  These analysis procedures then become reusable
and the basis for chains of more complicated analysis tasks.  Using such
chains, a user can volume render a set of halos based on their angular momentum
vectors, color particles by merger history, and even calculate disk inclination
angles and mass fluxes.

Once a region of the simulation is selected for analysis, \yt{} must process
the raw data fields themselves. Its model for handling this data and processing
fundamental data fields into new fields describing \emph{derived values} is
built on top of an object model with which we can build automatically recursive
field generators that depend on other fields.  All fields, including derived
fields, are allowed to be defined by either a component of a data file, or a
function that transforms one or more other fields.  This indirection allows
multiple layers of definition to exist, encouraging the user to extend the
existing field set as needed, using Python functions as transformation
and mathematical operators.

\section{Methods of Interacting with \yt{}}\label{sec:interacting}

The primary interface to \yt{} is through a programmatic API.  Scripts are
written and then executed, either in serial or in parallel through a batch
queue.  Interactive helper functions, implemented using IPython, are also
provided for tab-completion, figure handling and so forth.  We provide a
command-line utility with many common functions: plotting, statistics, volume
rendering, halo finding, pastebinning, image uploading, bug reporting, and even
uploading a script to the yt Hub (\texttt{http://hub.yt-project.org/}) to share
with other users.  Recently, the ability to spawn a Google Maps-like interface
has been added, to allow interactive panning and zooming of multi-resolution
datasets from a web-browser.

\begin{figure*}[!t]
\begin{centering}
\includegraphics[width=0.96\textwidth]{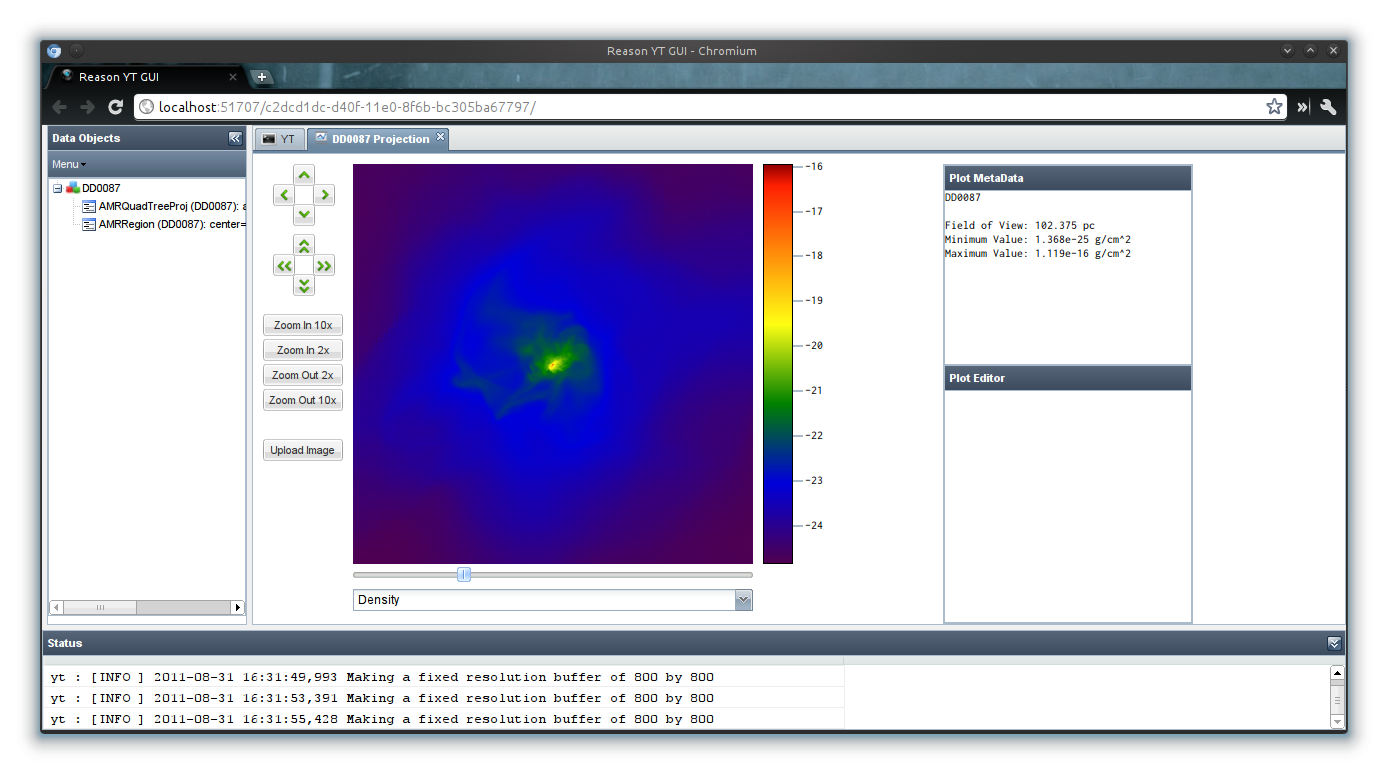}
\caption{A screenshot of the GUI ``Reason'' running in a local browser,
displaying data remotely processed and analyzed on a shared-user supercomputer
center.  This display shows a dynamically-created widget for exploring
simulation data.}
\label{fig:reason}
\end{centering}
\end{figure*}

The most recent version of yt (2.2) features a new GUI, entitled ``Reason,'' as
seen in Figure~\ref{fig:reason}.  At many supercomputing centers, toolkits for
constructing graphical user interfaces are not available or are extremely
difficult to build and install.  This greatly reduces the utility of creating a
traditional GUI.  To circumvent this limitation \yt{} provides a
fully-integrated GUI written in HTML and Javascript (ExtJS) and served by a
webserver (Bottlee and Rocket) running inside \yt{} itself.  Rather than a
large, bulky framework within which operations could be constructed and
executed, this GUI will presents a simple interactive interpreter that can be
accessed through a web browser.  This hosted interpreter dynamically creates
user interface widgets as well as enabling inline display of newly-created
images.  The primary user interface is a single-cell which can be submitted to
execute on the server; additional convenience features such as embedded IRC,
pastebin support, image upload and bug reporting are also included.  As this
GUI requires no client-side libraries or widgets, and as Python itself provides
all of the necessary tools on top of which this type of GUI could be built, we
believe this will be more maintainable and straightforward than a traditional
GUI.  A user creates a new server on-demand on a supercomputing center login
node and connects to it through an SSH tunnel from a local machine such as a
laptop.  Remote analysis and visualization are then guided and driven through
the locally-rendered web page, with results and images passed back
asynchronously and displayed inline in the same web page.  Future versions will
allow for parallel execution in batch queues and detachment and reattachment
operations.

\section{Visualization}\label{sec:visualization}

\yt{} provides methods for creating 2D and 3D visualizations of simulation
data.  The mechanisms for creating 2D visualizations have two primary
components: the data-handling portion and the figure creation or
``pixelization'' step.  The former is composed of a set of objects which
provide uniform access to 2D data objects, while the latter is a simple method
for making plots quickly, which can be wrapped into other convenience functions
(both created by \yt{} and external to \yt{}.)  The figure creation in \yt{} is
motivated by a desire for simplicity: rather than attempting to accommodate the
myriad use cases and user preferences, \yt{} seeks to provide a set of routines
that can be extended easily.  Users requiring complex figures for specific
publications can take the 2D image pixel buffers provided by \yt{} and feed
them to any plotting package, though \yt{} integrates most naturally with the
Matplotlib Python module \cite{matplotlib_paper}. Here, we first describe each
of the 2D pixalization mechanisms, and then the 3D volume rendering algorithms.
Futher information on the simple, built-in figure generation can be found in
the \yt{} documentation.

The simplest means of examining data is plotting grid-axis aligned slices
through the dataset.  This has several benefits - it is easy to calculate which
grids and which cells are required to be read off disk (and most data formats
allow for easy striding of data off disk, which reduces this operation's IO
overhead) and the process of stepping through a given dataset is relatively
easy to automate.

When handling astrophysical simulation data, one often wishes to examine either
the sum of values along a given sight-line or a weighted-average along a given
sight-line, in a \emph{projection}.  \yt{} provides an algorithm for generating
line integrals in an adaptive fashion, such that every returned
$(x_p,dx_p,y_p,dy_p,v)$ point does not contain data from any points where $dx <
dx_p$ or $dy < dy_p$; the alternative to this is a simple 2D image array of
fixed resolution perpendicular to the line of sight whose values are filled in
by all of the cells of the source object with overlapping domains.  But, by
providing this list of \emph{all} finest-resolution data points in a projected
domain, images of any field of view can be constructed essentially
instantaneously; conversely, however, the initial projection process takes
longer, for reasons described below.  We term the outputs of this  process
\emph{adaptive projections}.  For the Santa Fe Light Cone dataset
\cite{2007ApJ...671...27H}, to project the entire domain at the highest
resolution would normally require an image with $2^{30}$ values.  Utilizing
this adaptive projection method, we require less than $1\%$ of this amount of
image storage.

Direct ray casting through a volume enables the generation of new types of
visualizations and images describing a simulation.  \yt{} has the facility to
generate volume renderings by a direct ray casting method.  Currently the
implementation is implemented to run exclusively on the CPU, rather than faster
hardware-based rendering mechanisms, but this also allows for clearer
descriptions of the algorithms used for compositing, calculation of the
transfer function, and future advances in parallelization.  Furthermore, it
eases the task of informing volume renderings with other analysis results: for
instance, halo location, angular momentum, spectral energy distributions and
other derived or calculated information.  In \yt{}, volume rendering is exposed
through a ``Camera'' interface that allows for camera paths, zooms,
stereoscopic rendering and easier access to the underlying vector plane.
Transfer functions that can automatically sample colormaps as well as one that
provides off-axis line integrals are supplied, as well as a transfer function
whose colors correspond to Johnson filter-convolved Planck emission with
approximate scattering terms, as in \cite{vg06-kaehler}.  Utilizing the HEALpix
algorithm for equal latitudinal decomposition of a sphere
\cite{gorski-2005-622} \yt{} also provides the ability to render $4\pi$ images,
suitable both for creating outward-facing sky maps and planetarium images.

\begin{figure}[!t]
\begin{centering}
\includegraphics[width=0.46\textwidth]{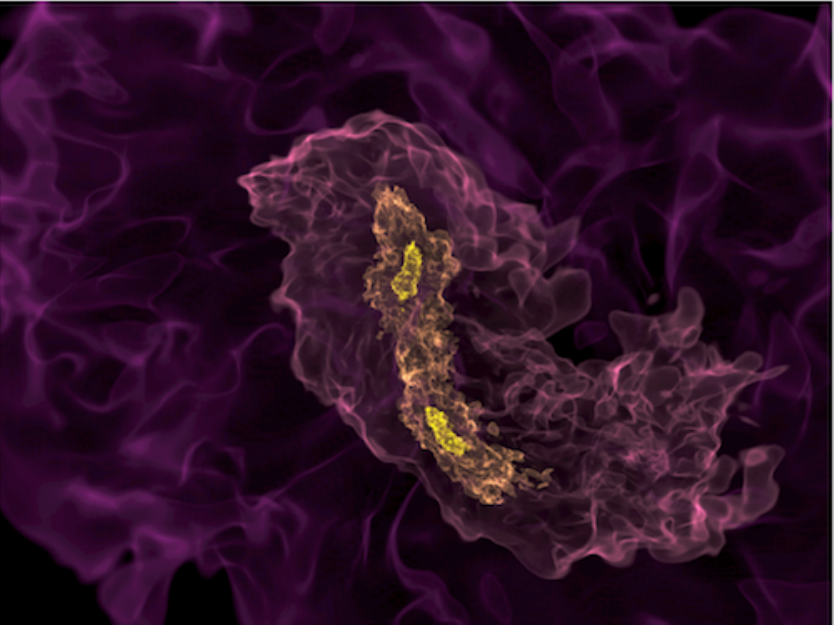}
\caption{A volume rendering of a metal-free star forming region that has fragmented
into two cores, each of which is likely to host a Population III star.  The field of
view is $2000~\mathrm{AU}$.  Isocontours were placed at $10^{-15}, 10^{-14},
10^{-13}$ and $10^{-12}~\mathrm{g}~\mathrm{cm}^{-3}$ \cite{2009Sci...325..601T}.}
\label{fig:vr_binary}
\end{centering}
\end{figure}

By allowing for detailed control over the specification of the transfer
function, viewing angle and generation of images, volume renderings that
contain a scientific narrative are easier to create.  For instance, in
Figure~\ref{fig:vr_binary} we have constructed a volume rendering of the
Population III star formation simulation described in
\cite{2009Sci...325..601T}, where a collapsing metal-free halo has been found
to fragment into two distinct clumps.  This volume rendering has been aligned
such that the normal vector to the image plane is aligned with the angular
momentum vector of the two-clump system.  Furthermore, the isocontours visible
in the image have been selected such that they coincide with transitions
between chemical states in the cloud.  Additional volume renderings based on
derived fields describing chemical and kinetic quantities could be constructed,
as well.

\section{Parallelism}\label{sec:parallelism}

As the capabilities of supercomputers grow, the size of datasets grows as well.
Most standalone codes are not parallelized; the process is time-consuming,
complicated, and error-prone.  Therefore, the disconnect between simulation
time and data analysis time has grown ever larger.  In order to meet these
changing needs, \yt{} has been modified to run in parallel on multiple
independent processing units on a single dataset.  Specifically, utilizing the
Message Passing Interface \cite{MPIStandard} via the \texttt{mpi4py} Python
module \cite{MPI4PY:PAPER1, MPI4PY:PAPER2}, a lightweight, NumPy-native wrapper
that enables natural access to the C-based routines for interprocess
communication, the code has been able to subdivide datasets into multiple
decomposed regions that can then be analyzed independently and joined to
provide a final result.  A primary goal of this process has been to preserve at
all times the API, such that the user can submit an unchanged serial script to
a batch processing queue, and the toolkit will recognize it is being run in
parallel and distribute tasks appropriately.

The tasks in \yt{} that require parallel analysis can be divided into two broad
categories: those tasks that act on data in an unordered, uncorrelated fashion
(such as weighted histograms, summations, and some bulk property calculation),
and those tasks that act on a decomposed domain (such as halo finding and
projection).  All objects and tasks that utilize parallel analysis exist as
subclasses of \texttt{ParallelAnalysisInterface}, which provides a number of
functions for load balancing, inter-process communication, domain decomposition
and parallel debugging.  Furthermore, \yt{} itself provides a very simple
parallel debugger based on the Python built-in \texttt{pdb} module.

To parallelize unordered analysis tasks, a set of convenience functions have
been implemented utilizing an initialize/finalize formalism; this abstracts the
entirety of the analysis task as a transaction.  Signaling the beginning and
end of the analysis transaction initiates several procedures, defined by the
analysis task itself, that handle the initialization of data objects and
variables and that combine information across processors.  These are abstracted
by an underlying parallelism library, which implements several different
methods useful for parallel analysis.  By this means, the intrusion of
parallel methods and algorithms into previously serial tasks is kept to a
minimum; invasive changes are typically not necessary to parallelize a task.
This transaction follows four steps.  First, the list of grids to process is
obtained.  This is followed by initialization of the parallelism on the data
object.  Each grid is then processed, and a finalize process is conducted on
the data object.  This is implemented through the Python iterator protocol; the
initialization of the iterator encompasses the first two steps and the
finalization of the iterator encompasses the final step.

Inside the grid selection routine, \yt{} decomposes the relevant set of grids
into chunks based on the organization of the datasets on disk.  Implementation
of the parallel analysis interface mandates that objects implement two
gatekeeper functions for both initialization and finalization of the parallel
process.  At the end of the finalization step, the object is expected to be
identical on all processors.  This enables scripts to be run identically in
parallel and in serial.  For unordered analysis, this process results in
close-to-ideal scaling with the number of processors.

In order to decompose a task across processors, a means of assigning grids to
processors is required.  For spatially oriented-tasks (such as projections)
this is simple and accomplished through the decomposition of some spatial
domain.  For unordered analysis tasks, the clear means by which grids can be
selected is through a minimization of file input overhead.  The process of
reading a single set of grid data from disk requires the opening of a file,
seeking to the position of the dataset in that file, the actual reading of the
data, and the file close operation.  For those data formats where multiple
grids are written to a single file, this process can be consolidated
substantially by performing multiple reads inside a single file once it has
been opened.  If we know the means by which the grids and fields are ordered on
disk, we can simplify the seeking requirements and instead read in large sweeps
across the disk.  By futher pre-allocating all necessary memory, this becomes a
single operation that can be accomplished in one ``sweep'' across each file.
By allocating as many grids from a single ``grid output'' file on a single
processor, this procedure can be used to minimize file overhead on each
processor.  Each of these techniques are implemented where possible. 

\begin{figure}[!t]
\centerline{\subfloat[Profiling]{\includegraphics[width=0.28\textwidth]{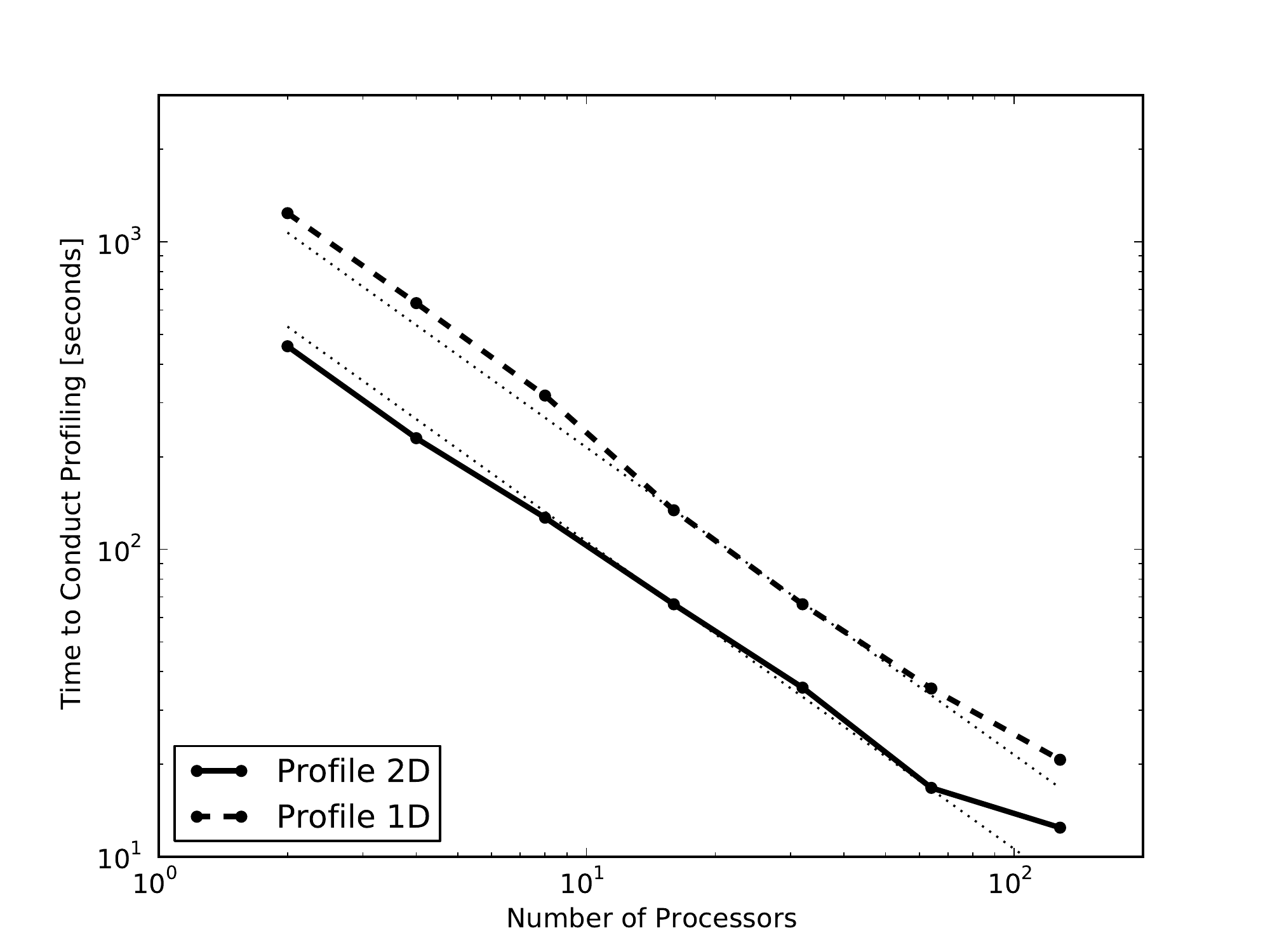}\label{fig:scaling_profiling}}
\hfil
\subfloat[Projecting]{\includegraphics[width=0.28\textwidth]{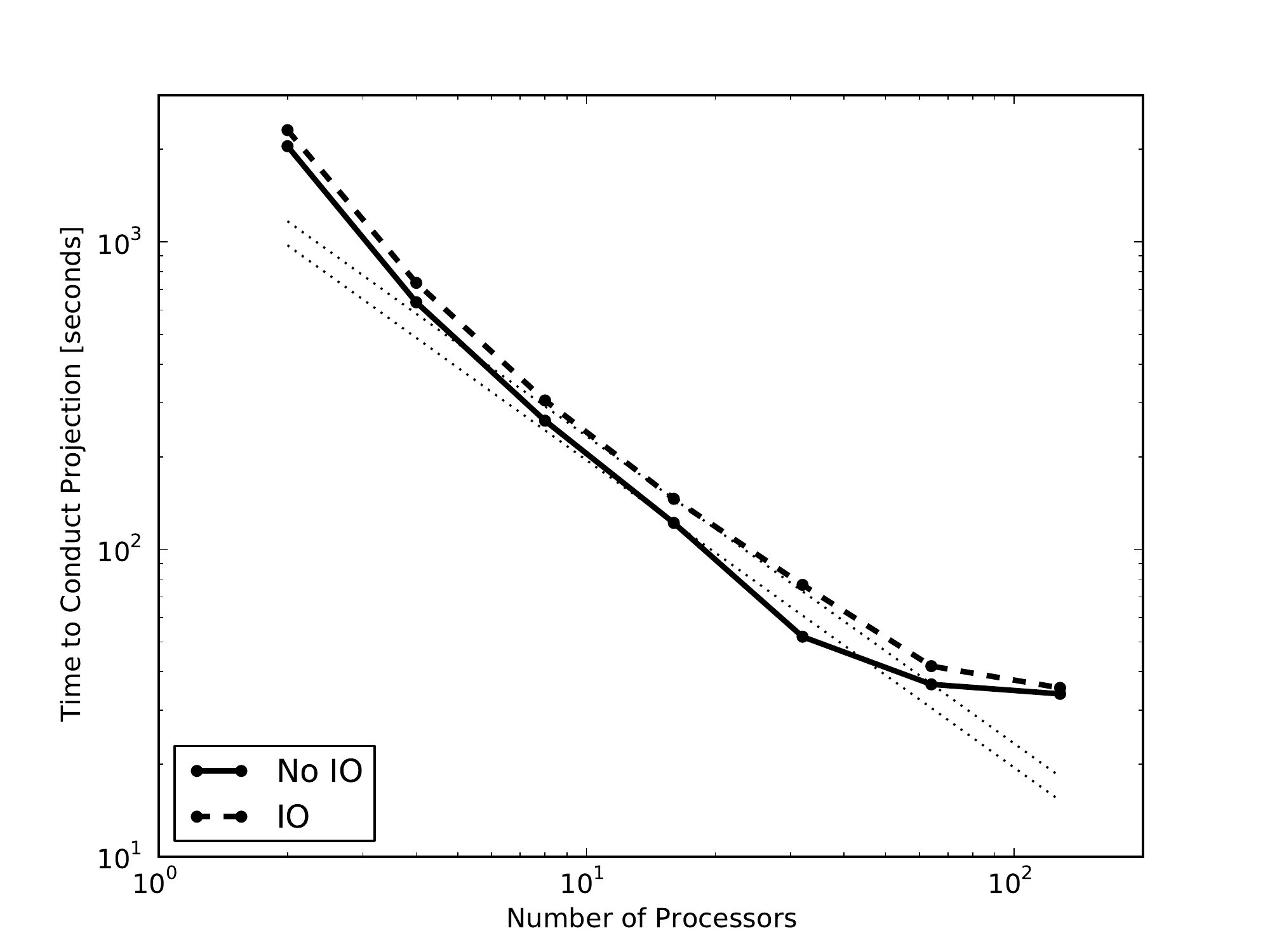}%
\label{fig:scaling_projection}}}
\caption{(left panel) Time taken for conducting 1- and 2-D profiles on the Santa Fe Light
Cone dataset at $z=0$ \cite{2007ApJ...671...27H}, a $512^3$ dataset with 6
levels of refinement (throughout the entire simulation domain) and a total of
$5.5\times10^8$ computational elements.  The overplotted thin solid lines
represent ideal scaling, as calibrated to the time taken by 16 processors.
(right panel) Time taken creating adaptive projections of the Santa Fe Light Cone
dataset at $z=0$ \cite{2007ApJ...671...27H}, a $512^3$ dataset with 6 levels
of refinement (throughout the entire simulation domain) and a total of
$5.5\times10^8$ computational elements.  In the case where IO was not
conducted, a field consisting uniformly of 1.0 everywhere was used as input.
The overplotted thin lines represent ideal scaling, as calibrated to the time
taken by 16 processors.}
\label{fig:scaling}
\end{figure}

In Figure~\ref{fig:scaling} (left panel) we show the results of a strong-scaling
study of conducting profiles of the final dataset from the Santa Fe Light Cone
\cite{2007ApJ...671...27H} project.  This dataset consists of
$5.5\times10^{8}$ computational elements.  The dashed black corresponds to
profiling in one dimension, and the solid line corresponds to profiling in two
dimensions.  Overplotted in thin solid lines are the ideal scaling
relationships, as calibrated to the time taken by 16 processors.  We see nearly
ideal strong scaling up to 128 processors, at which point overhead dominates;
we are essentially starving the processors of work at this scale.  The overall
time taken to conduct a profile is quite low, on one of the largest AMR
datasets in the published literature.  We note also that the substantial speed
difference between the two mechanisms of profiling, which is counter-intuitive,
is a result of a difference in implementation of the histogramming method; 1D
profiles use a pure-python solution to histogramming, whereas 2D profiles use a
hand-coded C routine for histogramming.  Future versions of \yt{} will
eliminate this bottleneck for 1D profiling and we expect to regain parity
between the two methods.

Several tasks in \yt{} are inherently spatial in nature, and thus must be
decomposed in a spatially-aware fashion.  MPI provides a means of decomposing
an arbitrary region across a given number of processors.  Through this method,
the \texttt{ParallelAnalysisInterface} provides mechanisms by which the domain
can be divided into an arbitrary number of subdomains, which are then realized
as individual data containers and independently processed.

For instance, because of the inherently spatial nature of the adaptive
projection algorithm implemented in \yt{}, parallelization requires
decomposition with respect to the image plane (however, future revisions of the
algorithm will allow for unordered grid projection.)  To project in parallel,
the computational domain is divided such that the image plane is distributed
equally among the processors; each component of the image plane is decomposed
into rectangular prisms (\texttt{AMRRegion} instances) along the entire line of
sight.  Each processor is allocated a rectangular prism of dimensions $(L_i,
L_j, L_d)$ where the axes have been rotated such that the line of sight of the
projection is the third dimension, $L_i \times L_j$ is constant across
processors, and $L_d$ is the entire computational domain along the axis of
projection.  Following the projection algorithm, each processor will then have
a final image plane set of points, as per usual: $$ (x_p, dx_p, y_p, dy_p, v)
$$ but subject to the constraints that all points are contained within the
rectangular prism as prescribed by the image plane decomposition.  At the end
of the projection step all processors join their image arrays, which are
guaranteed to contain only unique points.

In Figure~\ref{fig:scaling} (right panel) we show the results of a strong-scaling
study of adaptively projecting the same dataset as above.  The dashed line
represents a projection of the density field, whereas the solid line represents
projection in the absence of disk IO.  Clearly the algorithmic overhead
dominates the cost of disk IO, but we also see strong scaling between 4 and 64
processors; at 128 processors we see deviation from this.  The relatively early
termination of strong scaling (64 processors for this dataset, but we expect
this to be higher for larger datasets) as a result of algorithmic overhead is
one of the motivations behind future improvements to the projection algorithm.
However, from a pragmatic perspective, because \yt{} creates adaptive
projections, the time taken to project is a one-time investment and thus not a
rate-determining step for post-processed analysis.  For non-adaptive
projections, the process of handling all of the data, conducting parallel
reductions and outputting images must be undertaken for every chosen field of
view.

\section{Simulation Code Embedding}\label{sec:embedding}

\begin{figure}[!t]
\begin{centering}
\includegraphics[width=0.48\textwidth]{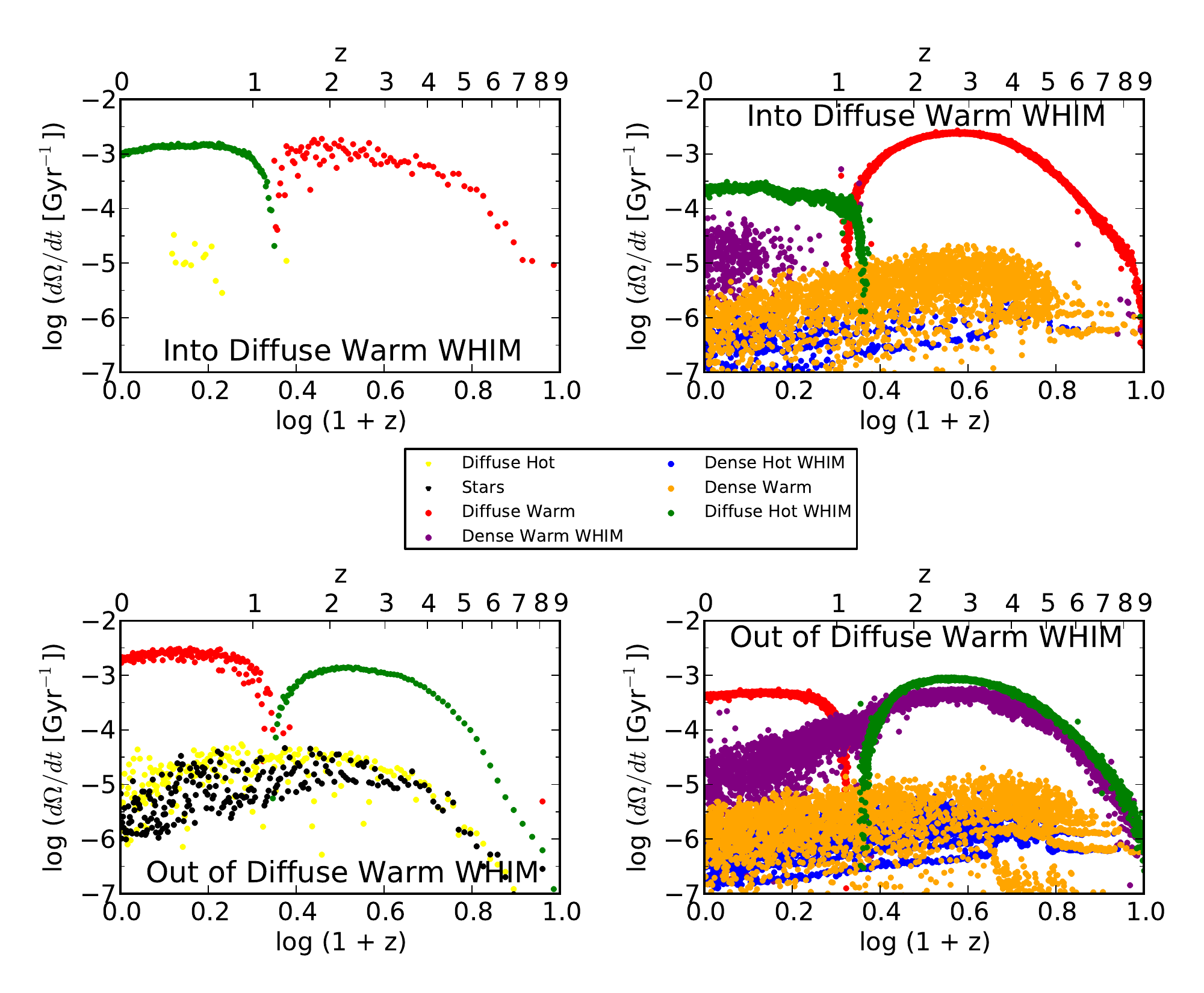}
\caption{The flux of matter into (top) and out of (bottom) the Diffuse
  Warm WHIM phase as a function of redshift, z, where z = 9
  corresponds to roughly 13.2 billion years in the past and z = 0 is
  today.  The Diffuse Warm WHIM phase is defined as matter within 
  the temperature range, $10^{5}$ K $< T < 10^{6}$ K, with densities
  less than 50 times the cosmic mean density.  The left panels show
  the results of this analysis performed using 252 simulation datasets
  written to  disk, while the right panels show the identical analysis
  performed at every single integration timestep with an in-situ
  instantiation of \yt{}.  Of note are three additional input and
  output phases identified in the in-situ analysis that were missed
  with the on-disk analysis.}
\label{fig:phase_flux}
\end{centering}
\end{figure}

An outstanding problem in the analysis of large scale data is that of
interfacing with disk storage; while data can be written to disk, read back,
and then analyzed in an arbitrary fashion, this process is not only slow but
requires substantial intermediate disk space for a substantial quantity of data
that will undergo severely reductionist analysis \cite{2007arXiv0705.1556N}.
To address this problem, the typical solution is to insert analysis code,
generation of derived quantities, images, and so forth, into the simulation
code.  However, the usual means of doing this is through either a substantial
hand-written framework that attempts to account for every analysis task, or a
limited framework that only handles very limited analysis tasks.  \yt{}
provides an explicit embedding API to enable in-line analysis.

By enabling in-line analysis, the relative quantity of analysis output is
substantially greater than that enabled by disk-mediated analysis; the cadence
of analysis tasks can be increased, leading to greater time-domain resolution.
Removing numerous large files dumped to disk as a prerequisite for conducting
analysis and generating visualization allows for a much more favorable ratio of
data to analyzed data.  For example, in a typical Population III star formation
simulation, such as in \cite{2009Sci...325..601T}, the size of the data dumps
can be as much as 10 gigabytes per timestep; however, the relative amount of
information that can be gleaned from these outputs is significantly smaller
\cite{2009Sci...325..601T}.  Using smaller data output mechanisms as well as
more clever streaming methods can improve this ratio; however, by enabling
in-line analysis, images of the evolution of a collapsing Population III halo
can be output at every single update of the hydrodynamical time, allowing for
true ``movies'' of star formation to be produced.  By allowing for the creation
and exporting of radial profiles and other analytical methods, this technique
opens up vast avenues for analysis while simulations are being conducted,
rather than afterward.

At the opposite end of the spectrum from simulations of Population III
star formation, which involve deep and complex adaptive-mesh
hierarchies, cosmological simulations of the evolution of large scale
structure, e.g. \cite{2011ApJ...731....6S},  are characterized by
static meshes that are much simpler but vastly greater in size.  The
largest simulations of \cite{2011ApJ...731....6S}, which had
1024$^{3}$ grid cells, required roughly 135 GB of disk space for each
dataset, of which 252 evenly spaced in time were written.  This
allowed for the study of time-dependent phenomenon with a resolution
of approximately 55 million years.  One of the primary goals of
\cite{2011ApJ...731....6S} was to understand the movement of matter
from one ``phase'' to another over cosmic time, where a phase is
determined simply by the density and temperature.  The flux of
material from one phase to another was calculated by comparing two
consecutive datasets written to disk and identifying grid cells in
each dataset in which the phase had changed.  An example of this is
shown in the left panels of Figure \ref{fig:phase_flux}, in which all of the inputs (top)
and outputs (bottom) into a single phase, the Diffuse Warm WHIM, are
plotted.  More recently, even larger simulations of this nature, with
1536$^{3}$ grid cells, were run with the identical analysis
performed.  However, instead of using consecutively written datadumps,
the analysis was performed during the simulation at every integration
timestep using the in-situ capabilities of \yt{} with grid data stored in
memory.  This allowed the phase flux analysis to be performed 5000
times instead of just 252, for a factor of 20 higher time resolution,
resulting in far greater insight into the phase evolution of matter,
as is illustrated in the right panels of Figure \ref{fig:phase_flux}.  The use of in-situ
analysis was all the more necessary in this simulation as each dataset
requires nearly 0.5 TB of disk space.  At this scale, performing this
analysis on datasets written to disk with the equivalent time
resolution would have require nearly 3 PB of storage space.

The Python/C API allows for passage of data in-memory to an instance of the
Python interpreter.  \yt{} has been instrumented such that it can be accessed
by an embedded Python interpreter inside a simulation code, such that one
interpreter instance exists for every MPI task.  \yt{} provides a clear API for
passing the necessary geometric information from the simulation code to the
analysis package.  By utilizing thin wrappers around the memory in which field
values and simulation data exist, the contents of the running simulation are
exposed to \yt{} and analysis can be conducted on them.  While this currently
works for many relatively simple tasks, it is not currently able to decompose
data spatially; as we are constrained by the parallel nature of most domain
decomposition algorithms, we attempt to avoid passing data between MPI tasks.
This means if a grid resides within MPI task 1, it will not be passed to MPI
task 2 during the analysis stage.  Currently this mechanism for inline analysis
has been exposed to Enzo simulations, and we hope to extend this in the future
to additional simulation codes.

Inline analysis will only become more important as simulations increase in size
and scope, and future development in \yt{} will make it easier, more robust,
and more memory efficient.  The primary mechanism by which \yt{} will be
embedded will change; future iterations of the inline analysis interface will
rely on communication between separate MPI jobs for simulation and analysis,
rather than an analysis task that shares memory space with the running
simulation code.  This mechanism will allow asynchronous analysis tasks to
be run, enabling the simulation to proceed while the user controls the data
that is examined.  Additionally, the method for interfacing \yt{} and
simulation codes will be provided as a single C++ library that can be
linked against, allowing it to be embedded by other developers.

\section{Future Directions}\label{sec:future_directions}

\subsection{Capabilities}

Development on \yt{} is driven by the pragmatic needs of working astrophysics
researchers.  Several clear goals need to be met in the future, particularly as
the size and scope of simulation data increases.  We also hope to work with
other research groups to add support for the visualization and analysis of
output from other popular astrophysics simulation codes such as ART, Gadget,
Pluto \cite{2007ApJS..170..228M}, and ZEUS-MP \cite{2006ApJS..165..188H}.

The most relevant improvement for very large simulation datasets is to improve
load balancing for parallel operations.  As noted above, for some operations
this can be accomplished by avoiding image-plane decomposition.  Several
efforts are underway to this end. Both the volume rendering and projection
algorithms load balance through decomposition of the image plane, which often
leads to poor work distribution.  These shortcomings are being addressed
algorithmically: adaptive projections will utilize a quad tree, enabling better
load balancing, and volume rendering will utilize a kD-tree approach combined
with intermediate image composition.  However, experimentation in quad tree
projection algorithms have indicated that the most rate-determining step
shifts, from IO resulting from poor load balancing to a time-consuming merger
step, wherein quad trees from different processors are merged.

However, an underlying problem with the parallelization as it stands is the
global state; each instance of a Python interpreter running \yt{} currently
runs in either ``parallel'' or ``serial'' mode.  Future versions of the \yt{}
parallel analysis interface will allow this to be toggled based on the task
under consideration, as well as more convenience functions for distributing
work tasks between processors--for instance, scatter/gather operations on
halos.  We intend to implement this on top of MPI, utilizing non-blocking
probes to function as a queueing and task distribution system.

Improvements to the communication mechanisms for parallel analysis in \yt{}
will be necessary as \emph{in situ} analysis grows more pervasive in large
calculations.  \emph{In situ} analysis is challenging as it must necessarily
proceed asynchronously with the simulation; this will require careful data
transport between \yt{} and the simulation code.  Abstracting and isolating the
engine that drives this communication will be necessary to enable \yt{} to be
embedded in simulation codes other than Enzo.  To this end, we have implemented
a ``Stream" frontend, suitable for supplying arbitrary data to \yt{}.  This can
function either as a remote endpoint for MPI intercommunicators, or as input
from ParaView, translating VTK objects obtained through ParaView's
Co-Processing functionality into \yt{} objects.

\subsection{Simulation Paradigm}

The process of instrumenting simulation codes for inline analysis provides
additional avenues for deeper control of the simulation code.  Typically, the
process of execution of a simulation involves an initialization step, a main
loop where modules that update the physical state of the simulation are
executed in sequence and the current time of the simulation is updated, and
then a finalization step, where memory is de-allocated, final outputs written
and the simulation is terminated.  The mechanism for calling physics modules is
either extremely specific to a given code or relatively cumbersome.  However,
once the simulation code's internal structures have been exposed to the broader
runtime environment, the conversion of this loop to a higher level language
becomes more practical and useful. This would enable rapid testing of
components such as load balancing schemes and physics modules.  The process of
modifying and debugging a code would be greatly simplified, and an interactive
iteration through the main loop would ease the process of inspecting and
debugging a simulation.  New users would be able to interactively step through
the physics modules, manually inspecting the updates to physical quantities and
learning how the simulation code behaves, rather than tediously examining and
recompiling.  Additionally, this provides the opportunity to interface more
readily with co-scheduled visualization tools through MPI Inter-communicators.

By abstracting the interface to underlying physics modules, individual physics
modules become trivially portable. In the future we intend to position \yt{} as
the outermost control structure for simulations of tera- and peta-scale
problems.  This will require substantial effort; the first steps in this will
be to identify an \textit{de facto}, rather than \textit{de jure} common API
for physics modules typically used in astrophysical simulation codes such as
Enzo.  We are developing wrappers for fundamental physics modules using Cython.
File system latency and read times become problematic even on the highest
performance Lustre file systems at processor counts in excess of $\sim1024$,
which requires the usage of a static-linking of utilized Python libraries as
well as the usage of zipfile-based module importing.  In addition to this, in
order to sidestep the issue of ABI incompatibilities with mismatched C++
compilers, we have eschewed the usage of Matplotlib for this purpose and
instead have developed a simplified PNG file writer that wraps both libpng and
freetype for rapid inspection of colorbars and plots.

Through this process, we intend to push forward in efforts to unify simulation
and analysis through high-level ``glue code'' such as \yt{} for the largest
scale simulations of star formation, galaxy formation, and the evolution of
galaxy clusters and the intergalactic medium.

\section{Conclusions}\label{sec:conclusions}

The \yt{} project is fully free and open source software, released under the
GNU General Public License, with no dependencies
on external code that is not also free and open source software.  The
development process occurs completely in the open at
\texttt{http://yt-project.org/}, with publicly-accessible source control
systems, bug tracking, mailing lists, and regression tests.  Building a
community of users has been a priority of the \yt{} development team, both to
encourage collaboration and to solicit contributions from new developers; both
the user and developer communities are highly distributed around the world.
\yt{} is developed using
Mercurial\footnote{\texttt{http://mercurial.selenic.com/}}, a distributed
version control system that enables local versioned development and encourages
users to make and contribute changes upstream.

Many of the operations conducted in \yt{}: fluid analysis, phase diagrams,
volume rendering, parallelism, and in situ analysis could feasibly be applied
to domains other than astrophysics.  We intend to generalize the underlying
code base such that it can be applied to many other data formats in
astrophysics, and ultimately we hope to provide these tools and techniques to
domains other than astrophysics.  Our first steps toward this, providing a
generic and arbitrary data loader, have shown that it is feasible.  Future
versions of \yt{} will generalize fields and particle handling, and should make
this process much easier.

The creation of a freely available, publicly inspectable platform for
simulation analysis allows the community to disentangle the coding process from
the scientific process.  Simultaneously, by making this platform public,
inspectable and freely available, it can be openly improved and verified.  The
availability and relatively approachable nature of \yt{}, in addition to the
inclusion of many simple analysis tasks, reduces the barrier to entry for young
scientists.  Rather than worrying about the differences between Enzo and FLASH
hierarchy formats, or row versus column ordering, or HDF4 versus HDF5 versus
unformatted fortran data formats, researchers can focus on understanding and
exploring their data.  More generally, however, by orienting the development of
an analysis framework as a community project, the fragmentation of methods and
mechanisms for astrophysical data analysis is greatly inhibited.  Future
generations of simulations and simulation codes will not only benefit from this
collaboration, but they will require it.

\section*{Acknowledgments}

M.J.T~was supported in this work by NSF CI TraCS fellowship award OCI-1048505.
B.D.S~acknowledges support by NASA grants ATFP NNX09-AD80G and NNZ07-AG77G and
NSF grants AST-0707474 and AST-0908199.  Both authors would like to thank the
users and developers of \yt{}, in particular Jeffrey S.~Oishi, Samuel
W.~Skillman, Stephen Skory, Cameron Hummels, and John H.~Wise.  We also thank
Tom Abel, Greg L.~Bryan, Berk Geveci, Charles Law, Michael L.~Norman, Brian
W.~O'Shea, Jorge Poco, and George Zagaris for thoughtful discussions.  \yt{}
has been supported directly or indirectly over the years by a number of
different funding agencies, including NSF, DOE and Academic institutions, and
we are grateful for their support.



%

\bibliographystyle{elsarticle-num}

\end{document}